\pdfoutput=1
\documentclass[10pt, journal, draftclsnofoot, onecolumn]{IEEEtran}
\usepackage{cite}
\usepackage{color}
\usepackage{graphicx}
\usepackage{tikz}
\usetikzlibrary{fit}
\usetikzlibrary{shapes}
\usetikzlibrary{intersections}
\usetikzlibrary{arrows.meta}
\pgfdeclarelayer{back}
\pgfsetlayers{back,main}
\usepackage{pgfplots}
\pgfplotsset{compat=1.9}
\usepackage{amsmath}
\usepackage{amsfonts}
\usepackage{booktabs}
\usepackage{array}
\usepackage{subcaption}
\usepackage{stfloats}
\fnbelowfloat
\usepackage{url}

\makeatletter
\pgfarrowsdeclare{X}{X}
{
  \pgfutil@tempdima=0.3pt%
  \advance\pgfutil@tempdima by.25\pgflinewidth%
  \pgfutil@tempdimb=5.5\pgfutil@tempdima\advance\pgfutil@tempdimb by.5\pgflinewidth%
  \pgfarrowsleftextend{+-\pgfutil@tempdimb}
  \pgfarrowsrightextend{0pt}
}
{
  \pgfutil@tempdima=0.3pt%
  \advance\pgfutil@tempdima by.25\pgflinewidth%
  \pgfsetdash{}{+0pt}
  \pgfsetroundcap
  \pgfsetmiterjoin
  \pgfpathmoveto{\pgfqpoint{-5.5\pgfutil@tempdima}{-6\pgfutil@tempdima}}
  \pgfpathlineto{\pgfqpoint{5.5\pgfutil@tempdima}{6\pgfutil@tempdima}}
  \pgfpathmoveto{\pgfqpoint{-5.5\pgfutil@tempdima}{6\pgfutil@tempdima}}
  \pgfpathlineto{\pgfqpoint{5.5\pgfutil@tempdima}{-6\pgfutil@tempdima}}
  \pgfusepathqstroke 
}
\makeatother

\begin{document}
\title{Network Slicing for Ultra-Reliable Low Latency Communication in Industry 4.0 Scenarios} 

\author{Anders~Ellersgaard~Kal{\o}r,
  Ren\'e Guillaume,
  Jimmy Jessen Nielsen,  
  Andreas Mueller,
  and  Petar Popovski%
  \thanks{Anders E. Kal{\o}r, Jimmy J. Nielsen, and Petar Popovski are
    with the Department of Electronic Systems, Aalborg University,
    9220 Aalborg, Denmark
    (e-mail: akalar12@student.aau.dk, \{jjn,petarp\}@es.aau.dk).}% <-this % stops a space
  \thanks{Ren\'e Guillaume and Andreas Mueller are with Corporate
    Sector Research and Advanced Engineering,
    Robert Bosch GmbH, 71272 Renningen, Germany
    (e-mail: \{Rene.Guillaume,Andreas.Mueller21\}@de.bosch.com).}}

\maketitle

\begin{abstract}
An important novelty of 5G is its role in transforming the industrial
production into Industry 4.0. Specifically, Ultra-Reliable Low Latency
Communications (URLLC) will, in many cases, enable replacement of
cables with wireless connections and bring freedom in designing and
operating interconnected machines, robots, and devices. However, not
all industrial links will be of URLLC type; e.g. some applications
will require high data rates. Furthermore, these industrial networks will be highly heterogeneous, featuring various communication technologies. We consider network slicing as a mechanism to handle the diverse set of requirements to the network. We present
methods for slicing deterministic and packet-switched industrial communication protocols at an abstraction level that is decoupled from the specific implementation of the underlying technologies. Finally, we show how network calculus can be used to assess the end-to-end properties of the network slices.
\end{abstract}

\IEEEpeerreviewmaketitle

\section{Introduction}
Industry 4.0 refers to the fourth industrial revolution
that transforms industrial manufacturing systems into cyber-physical
systems by introducing modern and emerging information and
communication technologies, such as 5G connectivity and cloud
computing~\cite{drath2014industrie,jazdi2014cyber}. Specifically, one
of the generic services in 5G termed 
Ultra-Reliable Low Latency Communications (URLLC) is poised to bring
wireless connections of unprecedented reliability, such as
$1-10^{-6}$~\cite{3gppTS22.261v16.0.0,schulz2017latency}. This will give rise to new
designs of machines and robots, released from the constraints imposed
by cabled connections and the need for physical attachment.
Nevertheless, not all connections will always require ultra-high
reliability. In some use cases high data rate may be required and in others, simultaneous support of many connections is required. In fact, the connected industry will feature connections of all three types envisioned in 5G: enhanced Mobile Broadband (eMBB), massive Machine-Type Communications
(mMTC) and URLLC~\cite{shafi20175g}.

\begin{figure}[!t]
\centering
\input{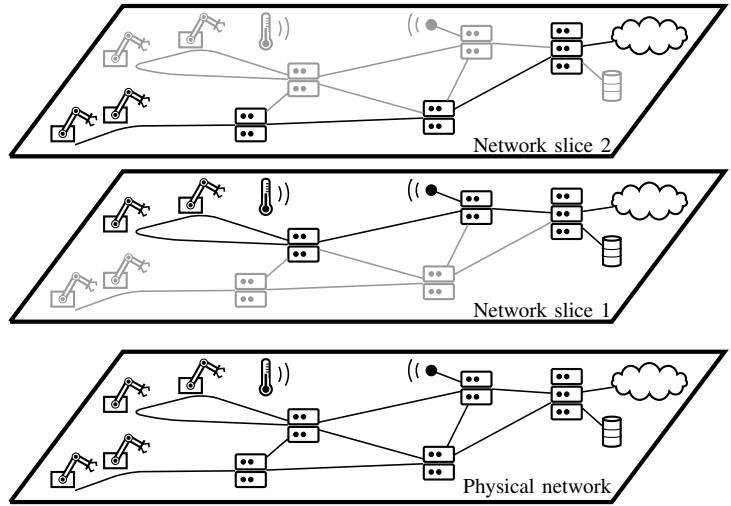}
\caption{Illustration of slicing a physical network into two logical
  sub-networks. Network slice 1 offers URLLC to an edge cache and the
  cloud, while network slice 2 provides an eMBB service to the Internet.}
\label{fig:networkslicing}
\end{figure}

Simultaneously satisfying diverse connectivity requirements within the same
system is challenging since the network cannot be optimized for a
specific type of service. A promising approach to handle this
problem is \emph{network slicing}, which refers to the process of
\emph{slicing} a physical network into logical sub-networks which are
each optimized for specific applications with certain
characteristics~\cite{alliance2016description}.
For instance, as illustrated in Fig.~\ref{fig:networkslicing}, one
network slice may offer URLLC based information
access by reserving communication and buffer resources along a path
from the end-user to an edge cache, or to a database in the cloud (network slice 1).
At the same time, another network slice in the same physical network may
offer an eMBB service between a robot and the Internet, e.g. to
allow for firmware updates (network slice 2).
Network slicing is enabled by recent network technologies such as
Software-Defined Networking (SDN) and Network Function Virtualization
(NFV) to decouple the network control plane from the data plane, and to
centralize the management of routing, queues, etc.

\subsection{Network Slicing for Industry 4.0}
Although encompassing eMBB, mMTC, and URLLC,
Industry 4.0 is characterized by its very strict latency and
reliability requirements. For instance, control systems and alarm
systems may require a delivery reliability in the order of $1-10^{-9}$
and end-to-end latencies in the range of
0.5--5~ms~\cite{schulz2017latency}.
As a result, network slicing for industrial networks poses several
challenges. First, constructing network slices
with strict end-to-end latency and reliability guarantees as required
by industrial applications is challenging due to the difficulties in
modeling and predicting queuing delays with high accuracy. Secondly,
industrial networks are often very heterogeneous, comprising
many specialized legacy protocols, which complicates accurate
end-to-end analysis~\cite{gaj2013computer}.
Guaranteeing low latency and high reliability has traditionally been
accomplished through the use of determinism and cyclic communication
with reserved resources and limited options for dynamic configuration.
This configuration is not ideal for Industry 4.0 where low latency
traffic not only may have to pass several links to reach the cloud,
but also is highly dynamic due to
the high mobility introduced by wireless technologies. Instead,
technologies with mechanisms for low-latency
communications may be more favourable, such as URLLC and Ethernet TSN which has
received much attention in the communities for industrial
communication~\cite{wollschlaeger2017future}.
However, it is unlikely that all protocols will be replaced by new technologies
at once, and network slices need to work across both new and
existing technologies. Therefore, network slicing must be studied and
resolved at an abstraction level which captures the main
characteristics of the protocols but is decoupled from the specific
implementations, such as legacy protocols, URLLC and Ethernet TSN.

In this article, we present methods for slicing industrial
communication protocols with focus on applications which require
strong reliability and latency guarantees analogous to those targeted by
URLLC. To this end, we investigate the utilization, reliability and isolation
trade-offs of the methods in an abstract setting which is independent
of the specific details of the protocols, and we demonstrate how
end-to-end properties of the proposed network slicing methods can be calculated
across communication technologies using network calculus, both for a
specific use case and in a general setting.
The remainder of the article is organized as follows.
Section~\ref{sec:netslicing} introduces
methods for slicing industrial networks. Section~\ref{sec:usecase}
describes a personalized medicine manufacturing system, which is used to
illustrate how end-to-end delivery reliability and latency bounds can be obtained.
Finally, the article is concluded in Section~\ref{sec:conclusion}.

\section{Network Slicing Methods}\label{sec:netslicing}
Industrial networks commonly follow a hierarchical structure as
illustrated in Fig.~\ref{fig:networkslicing}.
The individual devices such as actuators, sensors, etc. are connected
in a factory unit, and are
typically controlled by a master device in a master/slave
configuration. The connection may be wired or wireless, or in a
combination where a small 5G base station is part of the factory
unit, e.g. if there is need for high synchronization between the
devices. The factory unit is usually based on a deterministic and
cyclic protocol, with resources reserved to
the individual devices in each cycle. The cycle times may vary from
sub-millisecond to several milliseconds depending on the system.
The master devices of the individual factory units are connected to a
factory-wide network, which may also be connected to an external infrastructure
such as the Internet. The factory network is typically
based on switched protocols such as regular Ethernet or Ethernet TSN
and possibly TCP/IP. It includes general
purpose hardware and cloud computing resources which can be used by
the master devices, or even by components in a factory unit, and may comprise one or more 5G base
stations, which provide wireless connectivity to devices in the
factory.

We now describe slicing methods for cyclic protocols within factory
units, followed by a discussion and analysis of network slicing in
switched networks at the factory-wide network.

\subsection{Factory Units}
As a factory unit, we consider a single master/slave
network with a fixed cycle time. Each cycle contains a number of
resources (bytes), which are each allocated to a specific application
running on a certain device. The allocation is fixed and cannot change
during operation. We consider a network comprising one deterministic application
which transmits in every cycle (e.g. sensor readings for closed-loop
control), and $K$ stochastic applications which transmit frames
randomly (e.g. sensor alarms). The deterministic application transmits
$R_{\text{d}}$ frames of size $N_{\text{d}}$ in every cycle, while
the number of frames transmitted by stochastic application $k$ is
denoted by $R_{k}$, and of fixed size $N_{k}$.

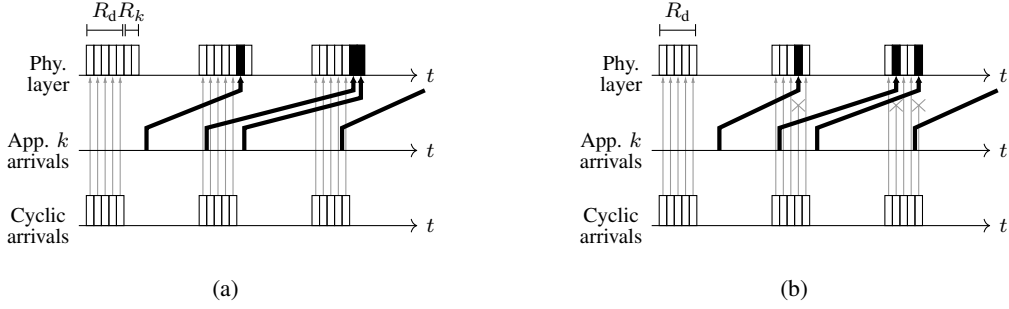
\begin{figure}[!t]
  \centering
  \begin{subfigure}[t]{0.44\textwidth}
    \centering
    \begin{tikzpicture}[font=\footnotesize]
  \linespread{0.9}
  \node[left,align=center] at (0,2cm) {Phy.\\layer};
  \draw[->] (0,2cm) -- ++(4.5cm,0) node[right] {$t$};
  \foreach \x in {0,...,2} {
    \foreach \y in {0,...,6} {
      \draw (0.1cm+\x*1.5cm+\y*0.1cm,2cm) rectangle ++(0.1cm,0.4cm);
    }
  }
  \draw[fill=black] (0.1cm+1.5cm+5*0.1cm,2cm) rectangle ++(0.1cm,0.4cm);
  \draw[fill=black] (0.1cm+2*1.5cm+5*0.1cm,2cm) rectangle ++(0.1cm,0.4cm);
  \draw[fill=black] (0.1cm+2*1.5cm+6*0.1cm,2cm) rectangle ++(0.1cm,0.4cm);

  \foreach \x in {0,...,2} {
    \foreach \y in {0,...,4} {
      \draw[black!40,-{Latex[length=1mm,width=0.5mm]}] (0.1cm+\x*1.5cm+\y*0.1cm+0.05cm,0.4cm) -- ++(0,1.6cm);
    }
  }

  \draw[|-|] (0.1cm,2.6cm) -- +(0.49cm,0) node[midway,above] {$R_{\text{d}}$};
  \draw[|-|] (0.61cm,2.6cm) -- +(0.19cm,0) node[midway,above] {$R_k$};
  
  %\draw[ultra thick,-{Latex[length=1mm,width=0.5mm]}] (0.9cm,1.4cm) -- (0.1cm+1*1.5cm+3*0.1cm+0.05cm,1.7cm);

  \node[left,align=center] at (0,1cm) {App. $k$\\arrivals};
  \draw[->] (0,1cm) -- ++(4.5cm,0) node[right] {$t$};
%  \foreach \x in {0.9,1.7,2.2,3.5} {
    \draw[ultra thick,-{Latex[length=1mm,width=1mm]}] (0.9cm,1cm) -- ++(0,0.3cm)
    -- (0.1cm+1*1.5cm+5*0.1cm+0.05cm,1.8cm) -- ++(0,0.2cm);
    \draw[ultra thick,-{Latex[length=1mm,width=1mm]}] (1.7cm,1cm) -- ++(0,0.3cm)
    -- (0.1cm+2*1.5cm+5*0.1cm+0.05cm,1.8cm) -- ++(0,0.2cm);
    \draw[ultra thick,-{Latex[length=1mm,width=1mm]}] (2.2cm,1cm) -- ++(0,0.3cm)
    -- (0.1cm+2*1.5cm+6*0.1cm+0.05cm,1.7cm) -- ++(0,0.3cm);
    \draw[ultra thick] (3.5cm,1cm) -- ++(0,0.3cm)
    -- (0.1cm+3*1.5cm,1.8cm);
%  }

    \node[left,align=center] at (0,0cm) {Cyclic\\arrivals};
  \draw[->] (0,0) -- ++(4.5cm,0) node[right] {$t$};
  \foreach \x in {0,...,2} {
    \foreach \y in {0,...,4} {
      \draw (0.1cm+\x*1.5cm+\y*0.1cm,0) rectangle ++(0.1cm,0.4cm);
    }
  }

\end{tikzpicture}
    \caption{}
    \label{fig:slicing_schemes_1}
  \end{subfigure}%
  \quad
  \begin{subfigure}[t]{0.44\textwidth}
    \centering
    \begin{tikzpicture}[font=\footnotesize]
  \linespread{0.9}
  \node[left,align=center] at (0,2cm) {Phy.\\layer};
  \draw[->] (0,2cm) -- ++(4.5cm,0) node[right] {$t$};
  \foreach \x in {0,...,2} {
    \foreach \y in {0,...,4} {
      \draw (0.1cm+\x*1.5cm+\y*0.1cm,2cm) rectangle ++(0.1cm,0.4cm);
    }
  }
  \draw[fill=black] (0.1cm+1.5cm+3*0.1cm,2cm) rectangle ++(0.1cm,0.4cm);
  \draw[fill=black] (0.1cm+2*1.5cm+1*0.1cm,2cm) rectangle ++(0.1cm,0.4cm);
  \draw[fill=black] (0.1cm+2*1.5cm+4*0.1cm,2cm) rectangle ++(0.1cm,0.4cm);

  \def\x{0}  
  \foreach \y in {0,...,4} {
    \draw[black!40,-{Latex[length=1mm,width=0.5mm]}] (0.1cm+\x*1.5cm+\y*0.1cm+0.05cm,0.4cm) -- ++(0,1.6cm);
  }
  \def\x{1}
  \foreach \y in {0,1,2,4} {
    \draw[black!40,-{Latex[length=1mm,width=0.5mm]}] (0.1cm+\x*1.5cm+\y*0.1cm+0.05cm,0.4cm) -- ++(0,1.6cm);
  }
  \foreach \y in {3} {
    \draw[black!40,-X] (0.1cm+\x*1.5cm+\y*0.1cm+0.05cm,0.4cm) -- ++(0,1.2cm);
  }
  \def\x{2}
  \foreach \y in {0,2,3} {
    \draw[black!40,-{Latex[length=1mm,width=0.5mm]}] (0.1cm+\x*1.5cm+\y*0.1cm+0.05cm,0.4cm) -- ++(0,1.6cm);
  }
  \foreach \y in {1,4} {
    \draw[black!40,-X] (0.1cm+\x*1.5cm+\y*0.1cm+0.05cm,0.4cm) -- ++(0,1.2cm);
  }
  
  \draw[|-|] (0.1cm,2.6cm) -- +(0.49cm,0) node[midway,above] {$R_{\text{d}}$};
  
  %\draw[ultra thick,-{Latex[length=1mm,width=0.5mm]}] (0.9cm,1.4cm) -- (0.1cm+1*1.5cm+3*0.1cm+0.05cm,1.7cm);

  \node[left,align=center] at (0,1cm) {App. $k$\\arrivals};
  \draw[->] (0,1cm) -- ++(4.5cm,0) node[right] {$t$};
%  \foreach \x in {0.9,1.7,2.2,3.5} {
    \draw[ultra thick,-{Latex[length=1mm,width=1mm]}] (0.9cm,1cm) -- ++(0,0.3cm)
    -- (0.1cm+1*1.5cm+3*0.1cm+0.05cm,1.8cm) -- ++(0,0.2cm);
    \draw[ultra thick,-{Latex[length=1mm,width=1mm]}] (1.7cm,1cm) -- ++(0,0.3cm)
    -- (0.1cm+2*1.5cm+1*0.1cm+0.05cm,1.8cm) -- ++(0,0.2cm);
    \draw[ultra thick,-{Latex[length=1mm,width=1mm]}] (2.2cm,1cm) -- ++(0,0.3cm)
    -- (0.1cm+2*1.5cm+4*0.1cm+0.05cm,1.8cm) -- ++(0,0.2cm);
    \draw[ultra thick] (3.5cm,1cm) -- ++(0,0.3cm)
    -- (0.1cm+3*1.5cm,1.8cm);
%  }

    \node[left,align=center] at (0,0cm) {Cyclic\\arrivals};
  \draw[->] (0,0) -- ++(4.5cm,0) node[right] {$t$};
  \foreach \x in {0,...,2} {
    \foreach \y in {0,...,4} {
      \draw (0.1cm+\x*1.5cm+\y*0.1cm,0) rectangle ++(0.1cm,0.4cm);
    }
  }

\end{tikzpicture}
    \caption{~}
    \label{fig:slicing_schemes_2}%
  \end{subfigure}%
  \caption{Two methods of cyclic resource multiplexing:
    \subref{fig:slicing_schemes_1}) multiplexing of reserved
    resources between gateway arrivals;
    \subref{fig:slicing_schemes_2}) multiplexing by overwriting
    random cyclic traffic.%
  }
\end{figure}

An obvious slicing scheme is to simply assign a number of resources in
each cycle to the individual applications based on the amount of data that they
transmit (Fig.~\ref{fig:slicing_schemes_1}). Suppose we allocate $N_k'$
bytes to stochastic application $k$.
Neglecting transmission and other error sources, and assuming that
excess frames are not buffered but dropped, the reliability of the
scheme is simply the probability that all
arriving frames can be transmitted, $\Pr(R_k N_k\le N_k')$.
Furthermore, assuming that $N_{\text{d}}'=R_{\text{d}}N_{\text{d}}$ bytes are
allocated to the deterministic application, it cannot fail due to
resource shortage. The proposed scheme provides a
high degree of isolation between applications and is simple
to analyze, but it results in a low resource utilization if data is not
transmitted in every cycle. This is particularly prominent for bursty
transmissions with high reliability requirements.
Although the utilization could be improved by introducing queuing to
the system, this complicates the analysis, in particular when the
arrival distribution take a more complex form, which makes it
difficult to provide end-to-end guarantees.

An alternative slicing scheme is where resources are allowed to be shared
between the $K$ stochastic applications. This results in an increased statistical
multiplexing gain due to the increased aggregate arrivals, and hence
an improved resource utilization.
Under this scheme, an allocation of $N'$ bytes are shared between $K$
applications, so that it fails when the aggregate arrival exceeds
$N'$, i.e. $\sum_k R_k N_k>N'$.
However, although the scheme increases the utilization, the gain comes
at the cost of reduced isolation due to multiplexing between the
applications. Specifically, the transmissions by one application
influence whether other applications can transmit. This may in
particular be problematic when there are uncertainties in traffic
models, due to its impact on the system reliability. Furthermore,
without introducing scheduling mechanisms, the scheme cannot take
diverse reliability requirements into account, e.g. through
application prioritizing. Therefore, it is most useful when multiple
applications transmit the same type of data, such as sensor readings.

Although multiplexing increases the utilization for high aggregate
arrival rates, it still achieves a low utilization for apps with low arrival
rates with high reliability requirements.
To improve the utilization for rare transmissions, we consider a
scheme where high-priority applications are allowed to overwrite
specific resources allocated to other applications (Fig.
\ref{fig:slicing_schemes_2}). This is also referred to as puncturing
in the context of 5G. For instance, a closed-loop control
system may obtain feedback from a
sensor in each cycle, but remain stable during short interruptions of the
feedback loop. Hence, the system provides a higher reliability than
needed. Suppose that each frame transmitted by a stochastic application overwrites a
random (uniformly distributed) frame from the deterministic
application. We assume that two application frames cannot overwrite
the same periodic frame, even if the sum of frame sizes is smaller
than the size of the periodic frame.
The reliability of the deterministic traffic is the probability that a
frame allocated to the deterministic application is not overwritten by
any of the application frames, while a stochastic application
transmission fails if the aggregate arrival $\sum_k R_{k}$ exceeds $R_{\text{d}}$.

\subsection{Factory-Wide Networks}
The factory-wide network is based on packet-switched
technologies where frames are queued at each link to increase the link
utilization. However, queuing introduces a random delay which depends
on the traffic that shares the link. In simple networks, the frames
may be processed as first-in-first-out, while complex networks may
apply various queue schedulers to control the flow and prioritize
certain types of traffic. A precondition for using switched networks
in Industry 4.0 is the ability to analyze the queuing delay of the
traffic with strict end-to-end latency requirements.
Several methods for queuing delay analysis exist, including queuing
theoretic approaches~\cite{kleinrock1976queueing} and stochastic
network calculus (SNC)~\cite{fidler2015guide}, which give
probabilistic results about the queuing delay, and deterministic
network calculus (DNC)~\cite{le2001network} which provides worst-case
latency bounds. While the probabilistic results from queuing
theory and SNC allow for exploiting the system requirements more
efficiently than DNC, the traffic arrival and server models are often
strongly restricted in order to keep the analysis tractable, which
limits the usefulness of the methods. In particular, they are not well
suited for industrial networks where the traffic is generated by a
mixture of periodic and stochastic sources, and where the network
requirements are too strict to allow for model approximations and
uncertainties. On the contrary, DNC allows for analyzing worst-case
latencies as long as the arrival processes are bounded by some
function. Since the network within a factory units has finite resources per
cycle, this provides a bound on the arrival processes. Therefore, we
focus on modeling the latency using DNC, although the other methods
could be applicable in some scenarios as well.
We omit a detailed presentation of DNC here, and instead refer
to~\cite{le2001network,bemten2016network} for a thorough treatment.

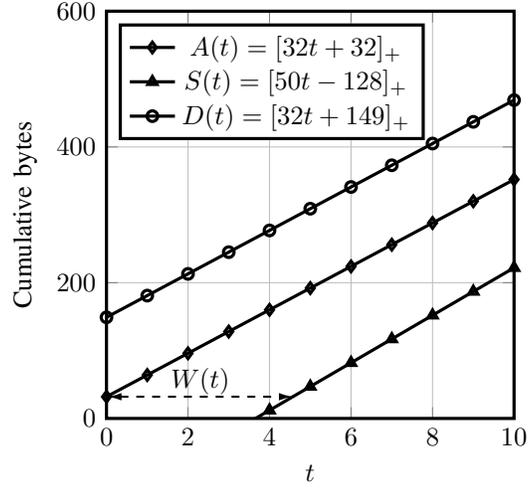
\begin{figure}[!t]
\centering
\begin{tikzpicture}
  \begin{axis}[
    ticks=both,
    grid=both,
    xlabel={$t$},
    ylabel={Cumulative bytes},
    xmin=0,
    ymin=0,
    ymax=600,
    xmax=10,
    mark size=2pt,
    very thick,
    legend pos=north west,
    width=7cm,
    height=7cm
    ]

    % Arrival
    % Arrival affine
    \addplot+[black,mark=diamond,domain=0:10,samples=11] {32*x+32};
    \addlegendentry{$A(t)=[32t+32]_+$};

    % Service affine
    \addplot+[black,mark=triangle,domain=0:10,samples=11] {35*x-128};
    \addlegendentry{$S(t)=[50t-128]_+$};
    
    % Arrival departure
    % sigma_a*t + rho_a + sigma_a*rho_s/sigma_s
    \addplot+[black,mark=o,domain=0:10,samples=11] {32*x+149};
    \addlegendentry{$D(t)=[32t+149]_+$};

    \draw[{Latex[length=2mm,width=1mm]}-{Latex[length=2mm,width=1mm]},thick,dashed](axis cs:0,32)--(axis cs:4.6,32)%
    node[midway,above=-0.1cm] {{$W(t)$}};
  \end{axis}
\end{tikzpicture}
\caption{Graphical representation of the quantities in DNC.}
\label{fig:dnc}
\end{figure}

The theory of DNC is based on the notion of arrival and service curves
which are functions that bound the cumulative number of bytes arriving
to and being served
by a queue. Although the theory of DNC is very general and results can
be obtained with many types of curves, the curves are often restricted
to affine bounds to simplify the analysis.
For example, the arrivals from an application that generates
a frame of size $N$ periodically every $M$ time units would be bounded by
the affine function $A(t)={[N/Mt+N]}_+$ where ${[x]}_+=\max(0,x)$.
Similarly, the service rate of a server, modeling the
serialization of frames, may be lower bounded by the affine service function.
Several results can be obtained from DNC, as exemplified in
Fig.~\ref{fig:dnc}. Here, $A(t)$ defines the affine bound on traffic
32 bytes arriving periodically in every cycle of duration $t$, and $S(t)$ is the
service curve defining the rate at which the arriving bytes are
served. From $A(t)$ and $S(t)$, one may obtain the waiting time bound
$W(t)$ and the departures from the queue, $D(t)$, which may in turn be
used as arrival curve
to the next queue in the path for end-to-end analysis.
Furthermore, through the notion of leftover service, DNC allows for
analyzing multiple queues with various scheduling policies such as
prioritization queuing. Leftover service refers to the
minimum service that is available to a queue after other queues have
been served. This is important in the context of industrial
applications, where scheduling, and in particular traffic
prioritization, is necessary to guarantee low end-to-end latencies
with high probability. Other quantities that can be obtained from DNC
include maximum queue size which can be used to dimension buffers.

\section{Case Study: Personalized Medicine Manufacturing}\label{sec:usecase}
This section introduces a simple personalized medicine manufacturing
system as a use case of Industry 4.0 to demonstrate how network
slicing can be used to handle diverse end-to-end network requirements.
Furthermore, the system will be used to study the trade-offs in the
slicing methods presented in previous section.
The system is derived to contain the main properties and realistic
requirements of an Industry 4.0 system, and is based on the potential
URLLC requirements defined by 3GPP~\cite{3gppTS22.261v16.0.0}. It is described in an
abstract way, which is independent of the specific communication
technologies used in the network, so that it can
represent both wired legacy protocols and URLLC technologies.

\begin{table*}
  \renewcommand{\arraystretch}{1.2}
  \caption{End-to-end requirements in the use case}
  \label{tab:usecase_req}
  \centering
  \begin{tabular}{llllllll}
    \toprule
    Application & Source & Dest. & Type & Mean period & Size & Latency req. & Reliability req. \\
    \midrule
    Control & Master & Robot & Periodic & 1 ms & 128 B & 1 ms & $1-10^{-4}$\\
    Control & Robot & Master & Periodic & 1 ms & 128 B & 1 ms & $1-10^{-4}$\\
    Patient info request & Master & Cloud & Periodic & 200 ms & 128 B & 10 ms & $1-10^{-4}$\\
    Patient info response & Cloud & Master & Periodic & 200 ms & 1024 B & 10 ms & $1-10^{-4}$\\
    Scale readings & Scale & Cloud & Periodic & 200 ms & 512 B & 100 ms & $1-10^{-6}$\\
    Sensor alarms & Receiver & Cloud & Poisson & 60 s & 32 B & 5 ms & $1-10^{-6}$\\
    HMI stream & Cloud & HMI & Periodic & 20 ms & 20 kB & 20 ms & $1-10^{-2}$\\
    \bottomrule
  \end{tabular}
\end{table*}
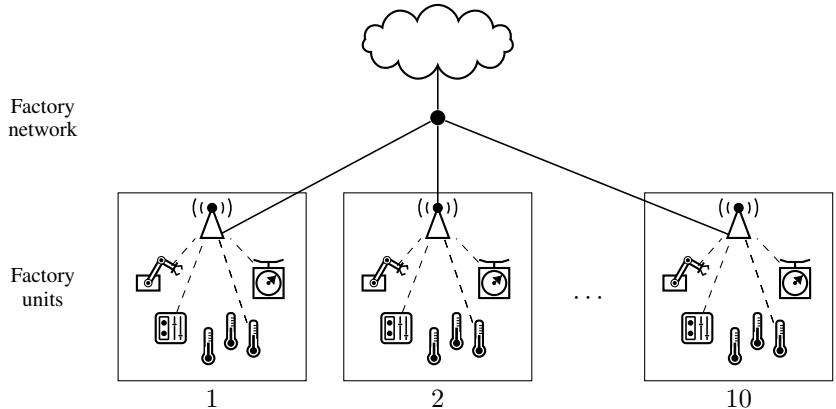
\begin{figure}[!t]
\centering
\begin{tikzpicture}
  \tikzset{
    pics/robot/.style args={#1:#2,#3:#4}{%
      code={
        \draw[fill=white,thick] (-0.15cm,-0.1cm) rectangle (0.15cm,0.1cm);
        \begin{scope}
          \draw[double distance=0.03cm, thick] (0,0)--(#1:#2);
          \draw[fill=white] (0,0) circle (0.05cm);
          \draw[fill=black] (0,0) circle (0.02cm);
        \end{scope}
        \begin{scope}[shift=(#1:#2)]
          \draw[double distance=0.03cm, thick] (0,0)--(#3:#4);
          \draw[fill=white] (0,0) circle (0.05cm);
          \draw[fill=black] (0,0) circle (0.02cm);
          \begin{scope}[shift=(#3:#4),rotate=#3]
            \draw[thick] (0,-0.08cm) -- (0,0.08cm);
            \draw[semithick] (0,0.02cm) -- (0.07cm,0.07cm) -- (0.1cm,0.03cm);
            \draw[semithick] (0,-0.02cm) -- (0.07cm,-0.07cm) -- (0.1cm,-0.03cm);
            \coordinate (claw) at (0.1cm,0.08cm);
          \end{scope}
        \end{scope}
        \coordinate (start) at (-0.15cm,-0.1cm);
        \node(-c) [fit = (start) (claw), inner sep = 0pt] {};
      }
    },
    pics/robot/.default={60:0.35cm,-30:0.2cm},
    pics/bs/.style={%
      code={
        \draw[thick,fill=white] (0,0) -- ++(-0.15cm,-0.4cm) -- ++(0.3cm,0) --
        cycle;
        \draw[fill] (0,0) circle (0.06cm);
        \draw[semithick] ([shift=(150:0.15cm)]0,0) arc (150:210:0.15cm);
        \draw[semithick] ([shift=(150:0.25cm)]0,0) arc (150:210:0.25cm);
        \draw[semithick] ([shift=(150:-0.15cm)]0,0) arc (150:210:-0.15cm);
        \draw[semithick] ([shift=(150:-0.25cm)]0,0) arc (150:210:-0.25cm);
        \coordinate (start) at (-0.25cm,-0.4cm);
        \coordinate (end) at (0.25cm,0.1cm);
        \node(-c) [fit = (start) (end), inner sep = 0pt] {};
      }
    },
    pics/wscale/.style={%
      code={
        \draw[thick,fill=white] (-0.2cm,-0.2cm) rectangle (0.2cm,0.2cm);
        \draw[thick,fill=white] (0,0) circle (0.16cm);
        \draw[fill=black] (0,0) circle (0.02cm);
        \draw[-{Latex[length=0.12cm,width=0.15cm]}] (0,0) -- (0.11cm,0.11cm);
        \draw[thick] (-0.2cm,0.31cm) -- (-0.1cm,0.28cm) -- (0.1cm,0.28cm) -- (0.2cm,0.31cm);
        \draw[thick] (0cm,0.2cm) -- (0cm,0.28cm);
        \coordinate (start) at (-0.2cm,-0.2cm);
        \coordinate (end) at (0.2cm,0.31cm);
        \node(-c) [fit = (start) (end), inner sep = 0pt] {};
      }
    },
    pics/ui/.style={%
      code={
        \draw[thick,rounded corners=0.05cm,fill=white] (0,0) rectangle +(0.4cm,0.4cm);
        \draw[semithick,rounded corners=0.01cm] (0.05cm,0.05cm) rectangle +(0.12cm,0.3cm);
        \draw[fill] (.11cm, 0.13cm) circle (0.03cm);
        \draw[fill] (.11cm, 0.27cm) circle (0.03cm);
        \draw[semithick,line cap=round] (0.25cm,0.05cm) -- +(0,0.3cm);
        \draw[white,thick,line cap=round] (0.23cm,0.25cm) -- +(0.04cm,0);
        \draw[semithick,line cap=round] (0.23cm,0.25cm) -- +(0.04cm,0);
        \draw[semithick,line cap=round] (0.33cm,0.05cm) -- +(0,0.3cm);
        \draw[white,thick,line cap=round] (0.31cm,0.1cm) -- +(0.04cm,0);
        \draw[semithick,line cap=round] (0.31cm,0.1cm) -- +(0.04cm,0);
        \coordinate (start) at (0,0);
        \coordinate (end) at (0.4cm,0.4cm);
        \node(-c) [fit = (start) (end), inner sep = 0pt] {};
      }
    },
    pics/thermometer/.style args={#1}{%
      code={
          \draw[thick,rounded corners=0.04cm] (-0.05cm,0) rectangle (0.05cm,0.4cm);
          \draw[thick,fill=white] (0,0) circle (0.08cm);
          \draw[draw=none,fill=white] (-0.036cm,0) rectangle (0.036cm,0.15cm);
          \draw[fill] (0,0) circle (0.04cm);
          \draw[fill] (-0.01cm,0) rectangle (0.01cm,0.2cm);
          \foreach \a in {0,...,3} {
            \draw (0,0.23cm+\a*0.035cm) -- +(0.04cm,0);
          }
          \coordinate (start) at (-0.08cm,-0.08cm);
          \coordinate (end) at (0.05cm,0.4cm);
          \node(-c) [fit = (start) (end), inner sep = 0pt] {};
      }
    },
    pics/thermometer/.default={}
  }

  \begin{scope}
    \draw (0cm,-0.5cm) rectangle (2.5cm,2cm);
    \pic[draw] (robot) at (0.4cm,0.8cm) {robot};
    \pic[draw] (scale) at (2.0cm,0.8cm) {wscale};
    \pic[draw] (ui) at (0.5cm,0.0cm) {ui};
    \pic[draw] (th) at (1.2cm,-0.2cm) {thermometer};
    \pic[draw] (th) at (1.5cm,0.0cm) {thermometer};
    \pic[draw] (th) at (1.8cm,-0.1cm) {thermometer};
    \begin{pgfonlayer}{main}
      \pic[draw] (bs1) at (1.25cm,1.8cm) {bs};
    \end{pgfonlayer}
    \draw[dashed] (robot-c) -- (bs1-c);
    \draw[dashed] (scale-c) -- (bs1-c);
    \draw[dashed] (ui-c) -- (bs1-c);
    \draw[dashed] (th-c) -- (bs1-c);
    \draw[dashed] (th-c) -- (bs1-c);
    \draw[dashed] (th-c) -- (bs1-c);
    \node[anchor=north] at (1.25cm,-0.5cm) {$1$};
  \end{scope}

  \begin{scope}[xshift=3cm]
    \draw (0cm,-0.5cm) rectangle (2.5cm,2cm);
    \pic[draw] (robot) at (0.4cm,0.8cm) {robot};
    \pic[draw] (scale) at (2.0cm,0.8cm) {wscale};
    \pic[draw] (ui) at (0.5cm,0.0cm) {ui};
    \pic[draw] (th) at (1.2cm,-0.2cm) {thermometer};
    \pic[draw] (th) at (1.5cm,0.0cm) {thermometer};
    \pic[draw] (th) at (1.8cm,-0.1cm) {thermometer};
    \begin{pgfonlayer}{main}
      \pic[draw] (bs2) at (1.25cm,1.8cm) {bs};
    \end{pgfonlayer}
    \draw[dashed] (robot-c) -- (bs2-c);
    \draw[dashed] (scale-c) -- (bs2-c);
    \draw[dashed] (ui-c) -- (bs2-c);
    \draw[dashed] (th-c) -- (bs2-c);
    \draw[dashed] (th-c) -- (bs2-c);
    \draw[dashed] (th-c) -- (bs2-c);
    \node[anchor=north] at (1.25cm,-0.5cm) {$2$};
  \end{scope}
  
  \begin{scope}[xshift=7cm]
    \draw (0cm,-0.5cm) rectangle (2.5cm,2cm);
    \pic[draw] (robot) at (0.4cm,0.8cm) {robot};
    \pic[draw] (scale) at (2.0cm,0.8cm) {wscale};
    \pic[draw] (ui) at (0.5cm,0.0cm) {ui};
    \pic[draw] (th) at (1.2cm,-0.2cm) {thermometer};
    \pic[draw] (th) at (1.5cm,0.0cm) {thermometer};
    \pic[draw] (th) at (1.8cm,-0.1cm) {thermometer};
    \begin{pgfonlayer}{main}
      \pic[draw] (bs3) at (1.25cm,1.8cm) {bs};
    \end{pgfonlayer}
    \draw[dashed] (robot-c) -- (bs3-c);
    \draw[dashed] (scale-c) -- (bs3-c);
    \draw[dashed] (ui-c) -- (bs3-c);
    \draw[dashed] (th-c) -- (bs3-c);
    \draw[dashed] (th-c) -- (bs3-c);
    \draw[dashed] (th-c) -- (bs3-c);
    \node[anchor=north] at (1.25cm,-0.5cm) {$10$};
  \end{scope}
    
  \node at (6.25cm,0.6cm) {$\mathbf{\cdots}$};

  \node[circle,fill=black,minimum size=0.2cm,inner sep=0pt] (nswitch) at (4.25cm,3cm) {};
  \begin{pgfonlayer}{back}
  \draw[semithick,black] (bs1-c.south) -- (nswitch);
  \draw[semithick,black] (bs2-c.south) -- (nswitch);
  \draw[semithick,black] (bs3-c.south) -- (nswitch);
  \end{pgfonlayer}
  
  \node[
  name path=cloud,
  cloud, cloud puffs=11,
  thick,
  minimum width=2cm, minimum height=1cm, draw,
  anchor=center
  ] (cloud) at (4.25cm,4cm) {};
  \draw[semithick,black] (nswitch) -- (cloud);

  \linespread{1}
  \node[align=center,font=\footnotesize] at (-1cm,0.75cm) {Factory\\units};
  \node[align=center,font=\footnotesize] at (-1cm,3cm) {Factory\\network};
\end{tikzpicture}
\caption{Personalized medicine manufacturing network consisting of
  $10$ factory units and a factory-wide network.}
\label{fig:usecase}
\end{figure}

The system consists of $10$ identical
master/slave factory units connected using an industrial cyclic communication
protocol. The master devices of each factory unit are physically
located at the base station, and connected to a
cloud through a hierarchical switched factory-wide network as depicted in
Fig.~\ref{fig:usecase}.
Each of the $10$ factory units controls a pipetting machine mounted on
a robotic arm, which dispenses a drug product into a container. The
type and amount of drug is determined based on patient information
obtained from a patient database in the cloud.
To validate the process the final product is weighed after the drug
has been dispensed, and the weight is stored in the cloud. Finally,
the entire process can be monitored by an operator using an
Human Machine Interface (HMI) which
is connected to the factory unit, and displays a video streamed from
the cloud. Furthermore, a number of sensors are located in
the unit to supervise the process, which may raise alarms in case of
failures. The requirements to the network are listed in Table~\ref{tab:usecase_req}.

The factory unit networks are based on a master/slave communication
protocol with a cycle time of $1~\text{ms}$. The factory-wide network is
based on switched 100~Mbit Ethernet, and the switches are equipped
with prioritization queues to each outgoing link. To simplify the
setting, we assume that each link only has two queues.
Furthermore, since we are mainly interested in the trade-offs in using
various slicing schemes, we consider a unified channel model where
the frame delivery reliability of the links in both the factory units
and the factory-wide network is $1-10^{-9}$.

\subsection{End-to-End QoS Analysis}\label{sec:endtoend}
There are numerous combinations of the network slicing schemes from
Section~\ref{sec:netslicing} that
may satisfy the application requirements in the medicine manufacturing
system, and a complete treatment is beyond the scope of this
article. Instead, we focus on a few applications and illustrate how
the proposed slicing schemes and DNC can be used to obtain end-to-end
results of the individual network slices, as well as to analyze the
interaction between the slices.
For simplicity, we ignore propagation delays and focus on queuing.
Furthermore, we ignore potential overhead added by protocol headers in
the network, and use milliseconds as the time unit.

\begin{figure}[!t]
\centering
\includegraphics{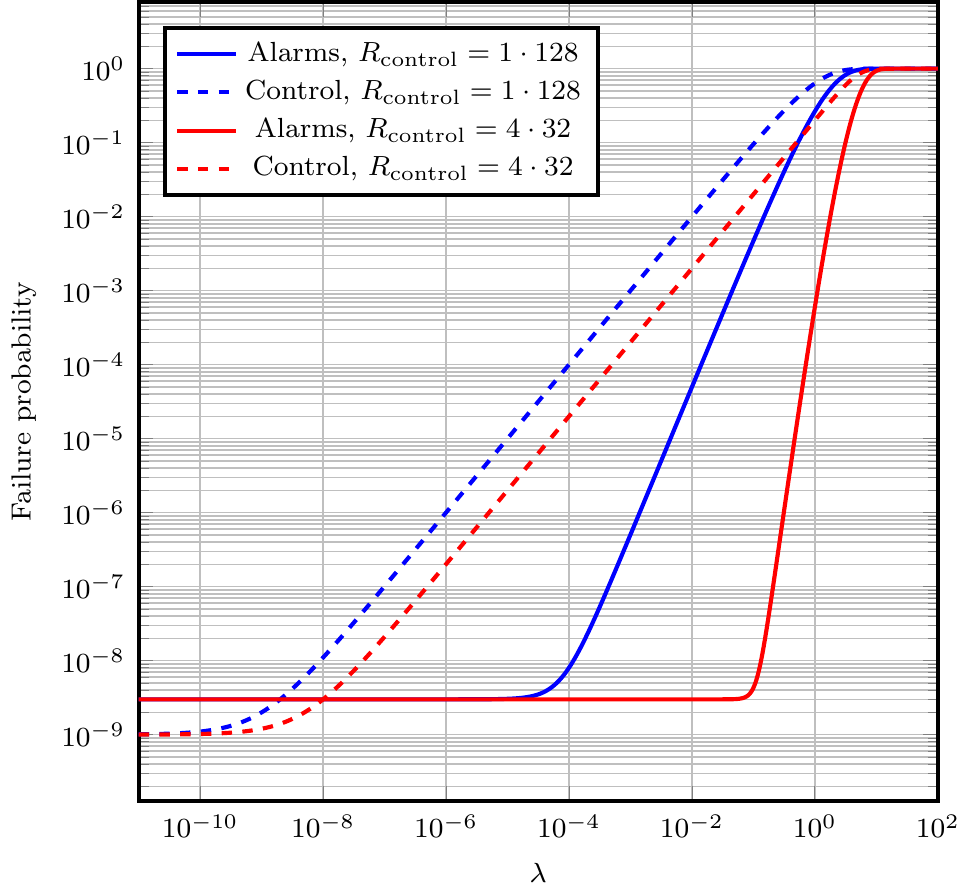}
\caption{End-to-end failure rate of control and sensor alarm traffic in the network slicing scheme based on overwriting for various alarm arrival rates.}
\label{fig:res_reliability}
\end{figure}

We first consider the resources for the sensor alarms. Since the
number of alarms in each cycle is random, we can either reserve a
fixed number of resources in each cycle, or we can allow alarms to
overwrite the cyclic control traffic, which has a lower reliability
requirement. Allocating a fixed number of resources results in a low
utilization, while overwriting control traffic introduces a decrease
in the reliability of the control traffic. Since the rate of alarms
is very low compared to the cycle time, the overwriting scheme is a
promising approach for this use case.
Figure~\ref{fig:res_reliability} shows the end-to-end frame failure
probabilities of the alarm and control traffic for a mean
number of alarm arrivals per cycle, $\lambda$. We consider the cases
where the control traffic comprises 1 and 4 frames,
$R_{\text{control}}=1\cdot 128$ and $R_{\text{control}}=4\cdot 32$.
At a low number of arrivals, the reliability approaches the
reliability of the links. Since there is only a single link
between the source and destination of the control traffic, compared to
three links for the alarms, its reliability is significantly higher.
As the number of arrivals increases, the reliability decreases for
both the control traffic and the alarms. The decrease in the control
traffic reliability is due to a higher probability of being
overwritten by an alarm, while the decrease in alarm reliability is
due to an increased probability of experiencing a shortage of
resources in a cycle. In the specific use case considered in this
article, the reliability requirement of the control traffic is
$1-10^{-6}$, which can be achieved up to an arrival rate of
$\lambda=4\cdot 10^{-6}$ for $R_{\text{control}}=4\cdot 32$.
Consequently, this is sufficient for the expected inter-arrival time
of 60~s ($\lambda\approx 1.7\cdot 10^{-4}$), and hence the overwriting
slicing scheme would be a reasonable choice. Furthermore, the
reliability of the alarm traffic at this point is very high since it
is unlikely that two alarms arrive in the same cycle, and since the
resources are used in all cycles, the utilization is 100 percent.
By comparison, if 32 bytes were allocated in each cycle only to the
sensor alarms, it would on average only be used once every
60 seconds, yielding a utilization of approximately 0.02 percent, and would
in addition occupy 32 bytes more of the frame than the overwriting scheme.

\begin{figure}[!t]
\centering
\includegraphics{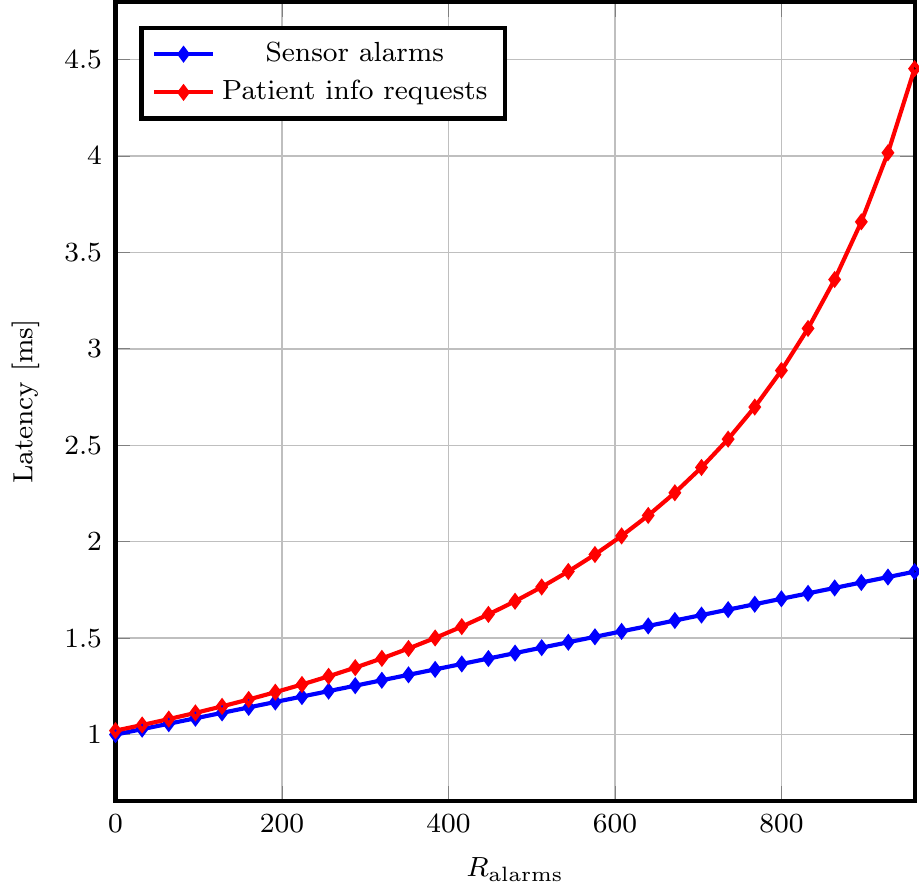}
\caption{End-to-end latency of sensor alarms and patient info requests for various alarm arrival rates.}
\label{fig:res_delay}
\end{figure}

A consequence of the prioritization queuing scheme in the factory
network is that the isolation between the queues is limited, since an
increase in the high-priority traffic also results in an increased
queuing delay of the traffic of lower priority.
Suppose now that we in the factory units decide to use the overwriting
scheme for the alarm traffic. Furthermore, assume that we give sensor
alarms high queue priority in the entire path from source to
destination, and that the periodic patient info requests are
given second priority. Obviously, the queuing delay that the patient
info requests experience depends on the number of sensor alarms.
The maximum number of bytes that can arrive to the factory network from the alarms
is enforced by the number of control frames that can be
overwritten. Specifically, in the cases considered above where
1 or 4 control frames are allocated, at most 1 or 4 alarm frames can
arrive to the factory network in each cycle (1~ms). Using this bound, we may use
DNC to obtain a bound on the total end-to-end latency experienced by
both the alarm frames and the patient info requests.
This is illustrated in Fig.~\ref{fig:res_delay} for various sensor
alarm arrival bounds, $R_{\text{alarms}}$. As $R_{\text{alarms}}$ approaches
$0$, the alarm latency approaches the cycle time of 1~ms.
For increasing $R_{\text{alarms}}$, the latency experienced by both the
sensor alarms and the patient info requests increases due to an
increased serialization time. Notice that despite being unchanged, the
patient info request latency increases with a larger slope than that
of the sensor alarms. This reflects the conservatism of the affine
bound, and is due to accumulation of low-priority frames in the time
where the high-priority traffic is served.
In a system with more queuing priorities, the
accumulation would occur at each prioritization queue all the
way to the queue with lowest priority. Although the latency is still
low in the shown scenario, it shows that enforcing a limit on the
number of bytes entering the factory-wide network is important to
maintain the required latency. This can either be done by exploiting
the reserved resources in the factory units as done here, or by inserting
traffic shapers, such as token buckets, into the network.

\section{Conclusion}\label{sec:conclusion}
5G, and particularly URLLC, will play an important role in
transforming industrial manufacturing systems into Industry 4.0.
Furthermore, a wide range of new applications will emerge due to the
increased connectivity, and they will have a diverse set
of requirements to the network,
ranging from ultra low-latency cyclic delivery guarantees to
best-effort and high data rates. This article
investigates network slicing as a way to handle this diverse set of
application requirements, with focus on URLLC.
We have presented methods for slicing both cyclic and switched
industrial protocols at an abstract level, and discussed their
trade-offs in utilization, reliability and isolation. Furthermore,
using a case study of an industrial medicine manufacturing system with
diverse network requirements, we have illustrated how deterministic network
calculus can be used for analyzing end-to-end latencies
of network slices comprising both deterministic and switched networks.

\section*{Acknowledgment}
The work by J. J. Nielsen and P. Popovski has partly been
supported by the European Research Council
(ERC Consolidator Grant nr. 648382 WILLOW),
and partly performed in the framework of the Horizon 2020 project ONE5G
(ICT-760809) receiving funds from the European Union.
The work by R. Guillaume and A. Mueller is supported by the
Federal Ministry for Education and Research (BMBF) within the project
Future Industrial Network Architecture (FIND) (16KIS0571).

\bibliographystyle{IEEEtran}
\bibliography{bibliography}
\end{document}